\documentclass{aa}  

\usepackage{textcomp}
\usepackage{graphicx}
\usepackage{txfonts}
\usepackage{longtable}
\usepackage{multirow}

\newcommand\nobr{\mbox{-}}
\newcommand\kms{km~s$^{-1}$}

\newcommand\micron{~\hbox{\textmu}m}
\begin{document}

   \title{ALMA observations of the kinematics and chemistry of disc formation}

   \author{Johan E. Lindberg
          \inst{1,2}
          \and
          Jes K. J{\o}rgensen\inst{2,1}
          \and
          Christian Brinch\inst{2,1}
          \and
          Troels~Haugb{\o}lle\inst{1,2}
          \and
          Edwin A. Bergin\inst{3}
          \and
          Daniel Harsono\inst{4,5}
          \and
          Magnus~V.~Persson\inst{4}
          \and
          Ruud Visser\inst{3}
          \and
          Satoshi Yamamoto\inst{6}
          }

   \institute{{Centre for Star and Planet Formation, Natural History Museum of Denmark, University of Copenhagen, {\O}ster Voldgade 5-7, DK\nobr1350 Copenhagen K, Denmark}\\
              \email{jolindbe@gmail.com}
         	\and
      {Niels Bohr Institute, University of Copenhagen, Juliane Maries Vej 30, DK\nobr2100 Copenhagen {\O}, Denmark}
        	\and
     {Department of Astronomy, University of Michigan, 500 Church Street, Ann Arbor, MI 48109-1042, USA}
     		\and
     {Leiden Observatory, Leiden University, P.O. Box 9513, NL-2300 RA Leiden, The Netherlands}
     		\and
     {SRON Netherlands Institute for Space Research, P.O. Box 800, NL-9700 AV, Groningen, The Netherlands}
     		\and
     {Department of Physics, The University of Tokyo, 7-3-1 Hongo, Bunkyo-ku, Tokyo, 113-0033, Japan}
         }

   \date{Received September 11, 2013; accepted May 4, 2014}


  \abstract
   {The R~CrA cloud hosts a handful of Class~0/I low-mass young stellar objects. The chemistry and physics at scales $>500$~AU in this cloud are dominated by the irradiation from the nearby Herbig~Be star R~CrA. The luminous large-scale emission makes it necessary to use high-resolution spectral imaging to study the chemistry and dynamics of the inner envelopes and discs of the protostars.}
   {We aim to better understand the structure of the inner regions of these protostars and, in particular, the interplay between the chemistry and the presence of discs.}
   {Using Atacama Large Millimeter/submillimeter Array (ALMA) high-resolution spectral imaging interferometry observations, we study the molecular line and dust continuum emission at submillimetre wavelengths.}
   {We detect dust continuum emission from four circumstellar discs around Class~0/I objects within the R~CrA cloud. Towards IRS7B we detect C$^{17}$O emission showing a rotation curve consistent with a Keplerian disc with a well-defined edge that gives a good estimate for the disc radius at 50~AU. We derive the central object mass to $2.3M_{\odot}$ and the disc mass to $0.024M_{\odot}$. The observations are also consistent with a model of material infalling under conservation of angular momentum; however, this model provides a worse fit to the data.
   We also report a likely detection of faint CH$_3$OH emission towards this point source, as well as more luminous CH$_3$OH emission in an outflow orthogonal to the major axis of the C$^{17}$O emission.}
   {The faint CH$_3$OH emission seen towards IRS7B can be explained by a flat density profile of the inner envelope caused by the disc with a radius $\lesssim50$~AU. We propose that the regions of the envelopes where complex organic molecules are present in Class~0/I young stellar objects can become quenched as the disc grows.}

   \maketitle
%

\section{Introduction}

The earliest deeply embedded protostellar stages (Class 0 and I)
represent an important period in the formation of solar-type stars and
their discs. In particular, the physical and chemical evolution during
these few hundred thousand years may represent the initial conditions
for solar system formation -- and serve as the link between the
environment in which the stars form and the outcome in terms of the
protoplanetary discs and their composition. With the high resolution and
sensitivity offered by the Atacama Large Millimeter/submillimeter
Array (ALMA) it is becoming possible to routinely image the innermost
regions of such systems and thus shed light on the distribution of gas
and dust there. In this paper, we present ALMA Early Science (Cycle 0)
observations of deeply embedded protostars in the Corona Australis
region -- in particular, those representing a deep search for the presence
of complex organic molecules and imaging of the dynamics in the
innermost regions around these sources.

One of the big unanswered questions about the formation of
circumstellar discs is exactly when those first appear during the
evolution of young stars. Previous continuum surveys point to the
existence of compact and dense regions of dust above the centrally
condensed envelopes around even very deeply embedded Class~0
protostars
\citep[e.g.,][]{harvey03,looney03,jorgensen04,jorgensen09,enoch11},
but the kinematics and thus the exact nature of these components remain
unclear: are they a representation of the earliest Keplerian discs,
do they instead represent earlier magnetic pseudo-discs or are they another
consequence of the protostellar core collapse? Recent observations of
one deeply embedded Class~0 protostar, L1527 \citep{tobin12b}, as well
as some Class~I young stars
(e.g., \citealt{brinch07,brinch13,lommen08,jorgensen09,harsono14}) show
evidence of Keplerian motions on $\sim 100$~AU scales, but in
general it remains challenging to disentangle these components from
the collapsing envelope \citep[see, e.g.,][]{tobin12a}.

The exact physical nature of the innermost regions of protostellar envelopes is strongly related to the chemistry
of these regions. One of the very exciting discoveries over the last
decade has been the detection of complex organic molecules in
the warm ($T > 100$~K) innermost regions of some deeply embedded
protostars \citep[sometimes referred to as ``hot cores'' or ``hot corinos''; see, e.g.,][]{bottinelli04,jorgensen05,ceccarelli07,herbst09} where the increase in temperature is thought to enhance grain mantle evaporation -- but any more
quantitative interpretation of these observations is strongly coupled
to the understanding of the envelope on these scales ($\lesssim 30$~AU). It also remains
a question whether such warm regions with complex organics are present
only in selected sources (possibly an indication of a particular
physical and chemical evolution of those) or whether the absence of complex organic
molecules from single-dish observations in some sources rather could
be a consequence of lack of sensitivity and/or unfavourable
physics. For example, if discs form early in the evolution of
protostars, they may dominate the mass budget in the innermost regions
of low-mass protostellar envelopes and only a small fraction of the
material there may have high enough temperatures for water and complex
organics to be in the gas phase in significant amounts to allow them
to be detectable.

We present ALMA observations of four young stellar objects (YSOs) in the R~CrA cloud (NGC~6729) within the Corona Australis star-forming region, one of the nearest star-forming regions, at a distance of 130~pc \citep{neuhauser08}. \citet{peterson11} found 116 YSO candidates in the Corona Australis star-forming region by multi-wavelength photometry analysis, the vast majority being Class~II or older, whereas the highest concentration of Class~I or younger YSOs was found in the R~CrA cloud. This cloud has its name from the Herbig~Be star R~CrA, which irradiates the protostellar envelopes, increasing the temperature of the molecular gas to $>30$~K on large ($>5000$~AU) scales \citep{vankempen09a,lindberg12}.

IRS7B is a low-mass ($L\approx4.6~L_{\odot}$) YSO in the R~CrA cloud located $45$\arcsec\ (6000~AU) southeast of the star R~CrA. Through infrared photometry, it was determined to be Class~I or younger by \citet{peterson11}. Through \textit{Herschel} far-infrared observations its bolometric temperature and $L_{\mathrm{bol}}/L_{\mathrm{submm}}$ ratio can be established. The bolometric temperature is 89~K, and the bolometric/submillimetre luminosity ratio is 48, which put it as a borderline Class~0/I source \citep{lindberg13}. To understand the physical stage of this source, other methods are required.

The remaining three YSOs covered by our ALMA observations -- CXO~34, IRS7A, and SMM~1C -- are also situated within the R~CrA cloud. These are all defined as Class~I or younger by \citet{peterson11}. The two sources IRS7A and SMM~1C cannot be spatially resolved by \textit{Herschel} PACS observations. Their combined spectral energy distribution (SED) corresponds, like in the case of IRS7B, to borderline Class~0/I sources. For a summary of previous (sub)millimetre observations of the region, refer to Sect.~1 of \citet{lindberg12}.

Sect.~2 presents the methods of observations and data reduction. Sect.~3 shows the observed spectra and continuum and spectral line maps. In Sect.~4 we study the kinematics of the molecular gas around IRS7B and describe the radiative transfer modelling of CH$_3$OH emission from IRS7B. Sect.~5 discusses the implications of the results and Sect.~6 summarises the conclusions of our study.

\section{Observations}

R~CrA IRS7B was observed on 4, 6, and 9 May 2012 in ALMA band~7 as part of ALMA Cycle~0 observations. The total integration time of the observations was 5.5~hours. During the observations, 18 antennas were part of the array, positioned in an extended configuration with baselines of 30--380~m (34--$430~\mathrm{k}\lambda$).

The complex gains were calibrated using the quasar J1924-292 located 10\degr\ from R~CrA IRS7B, which was observed every 15 minutes. The bandpass calibration was performed using the same quasar. The quasar flux was calibrated with observations of Neptune and Titan. 

The observations were performed in four spectral windows simultaneously (see Table~\ref{tab:spw} for details), each with 3840 channels with channel widths of 122~kHz ($\sim0.11$~km~s$^{-1}$). The typical rms in line-free channels is 9--10~mJy~beam$^{-1}$~channel$^{-1}$. A few hundred channels at some of the spectral window edges were flagged due to considerable noise. The line-free channels were used to record the continuum emission. The continuum data were self-calibrated to reduce side-lobes and increase the signal-to-noise ratio.

The spectral line data were cleaned using H{\"o}gbom PSFs in the minor cycles and Briggs weighting with a robustness parameter of 0.5 to acquire a balance between signal-to-noise and resolution performance. This resulted in a synthesised beam size of approximately $0\farcs42\times0\farcs37$ ($\mathrm{P.A.} \approx 72\degr$).

\begin{table}
\centering
\caption[]{Spectral windows of the observations.}
\label{tab:spw}
\begin{tabular}{r l r r r}
\noalign{\smallskip}
\hline
\hline
\noalign{\smallskip}
ID & Frequency range\tablefootmark{a} & rms\tablefootmark{b} & \multicolumn{2}{l}{Channel width} \\
& [GHz] & [mJy] & [kHz] & [km~s$^{-1}$] \\
\noalign{\smallskip}
\hline
\noalign{\smallskip}
0 & 336.955--337.425 & \phantom{0}9 & 122 & 0.109 \\
1 & 338.276--338.721 & \phantom{0}9 & 122 & 0.108 \\
2 & 349.344--349.814 & 10 & 122 & 0.105 \\
3 & 346.921--347.354 & 10 & 122 & 0.105\\
\noalign{\smallskip}
\hline
\end{tabular}
\tablefoot{
	\tablefoottext{a}{For emission at $\varv_{\mathrm{LSR}}=0$~km~s$^{-1}$. The IRS7B $\varv_{\mathrm{LSR}}\approx6$~km~s$^{-1}$ corresponds to frequencies $\sim7$~MHz higher.}
	\tablefoottext{b}{Typical rms per beam per channel around lines in this spectral window.}
     	}
\end{table}

\section{Results}

\subsection{Continuum emission}

\begin{figure*}[!tb]
    \centering
    \includegraphics{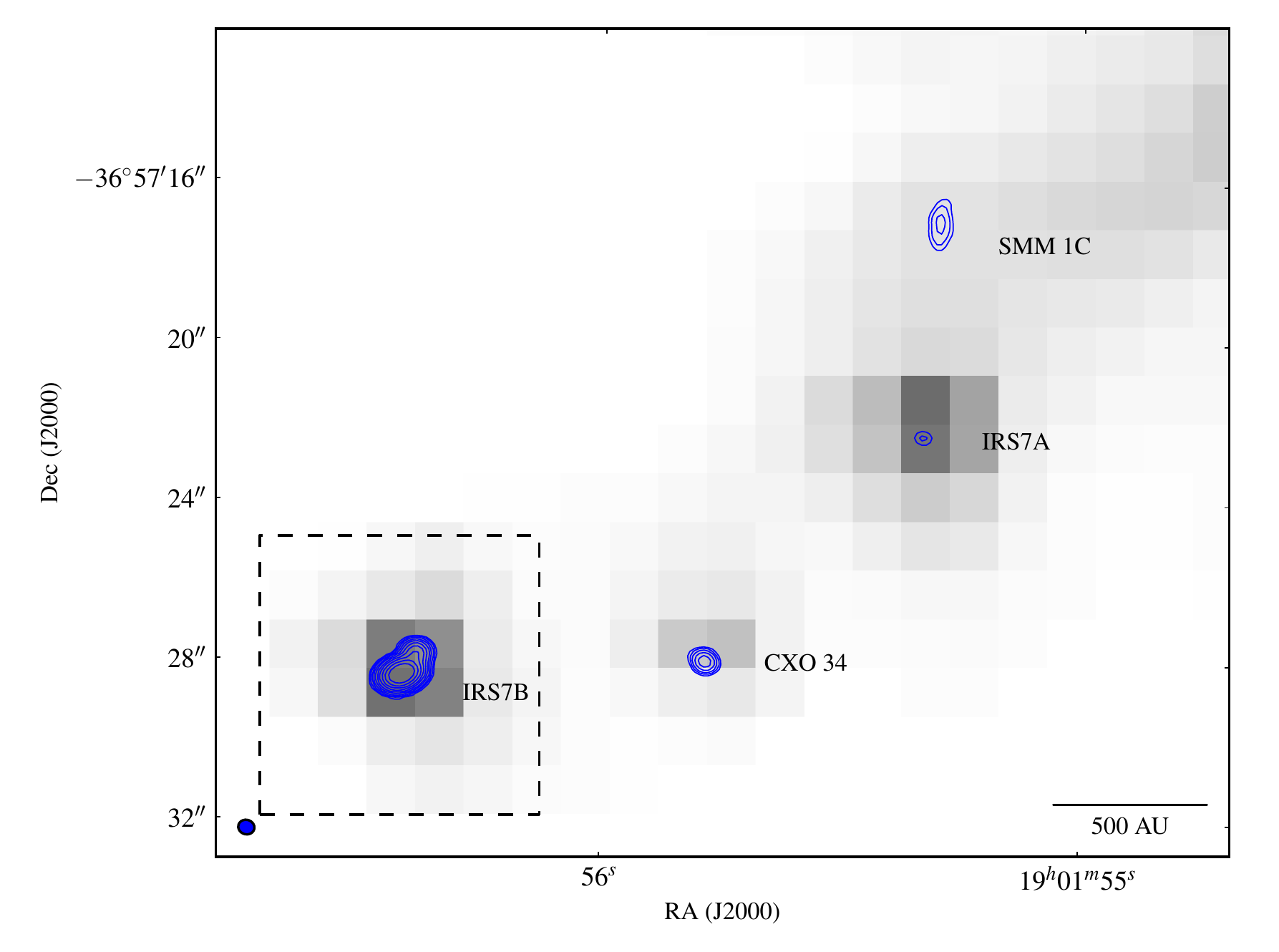} \\
    \caption{ALMA 342~GHz continuum (blue contours) overplotted on \textit{Spitzer} 4.5\micron\ image. The contours are logarithmically spaced, with the first contour at 2~mJy and the tenth and last contour at 200~mJy. The ALMA data are not primary-beam corrected, making the sources far from the phase centre at IRS7B appear fainter than they are. The dashed box shows the coverage of the moment maps in Fig.~\ref{fig:mom0a}.}
    \label{fig:sources_cont}
\end{figure*}

In the continuum emission, four point sources were identified. It was possible to match two of these with previous 1.3~mm SMA detections (IRS7B and SMM~1C). The remaining two sources (CXO34 and IRS7A), as well as IRS7B, are well-aligned with previously detected \textit{Spitzer}~4.5\micron\ point sources. Measured coordinates of the detected point sources are shown in Table~\ref{tab:pointsource} along with fitted continuum fluxes and derived disc masses. A continuum map can be found in Fig.~\ref{fig:sources_cont}.

\begin{table}[!tb]
\centering
\caption[]{Positions, continuum fluxes and disc masses of the detected point sources.}
\label{tab:pointsource}
\begin{tabular}{l l l l l}
\noalign{\smallskip}
\hline
\hline
\noalign{\smallskip}
Name & RA & Dec & Flux\tablefootmark{a} & $M_{\mathrm{disc}}$\tablefootmark{b} \\
& (J2000.0) & (J2000.0) & [mJy] & [$M_{\odot}$] \\
\noalign{\smallskip}
\hline
\noalign{\smallskip}
IRS7B & 19:01:56.42 & $-$36:57:28.4 & $432\pm\phantom{0}1$ & 0.024 \\
CXO~34 & 19:01:55.78 & $-$36:57:27.9 & $\phantom{0}46\pm\phantom{0}4$ & 0.003 \\
IRS7A & 19:01:55.33 & $-$36:57:22.4 & $\phantom{0}39\pm21$ & 0.002 \\
SMM~1C & 19:01:55.31 & $-$36:57:17.0 & $233\pm50$ & 0.013 \\
\noalign{\smallskip}
\hline
\end{tabular}
\tablefoot{
	\tablefoottext{a}{Primary-beam-corrected flux from a 2-D gaussian fit to the cleaned continuum image. The error reflects the accuracy in the gaussian fit, not the significance of the detection.}
	\tablefoottext{b}{Disc masses calculated using the method of \citet{jorgensen07}, which assumes optically thin emission and a dust temperature of 30~K.}
     	}
\end{table}

\subsection{Spectral line emission}

\begin{figure*}[!htb]
    \centering
    $\begin{array}{c@{\hspace{0.0cm}}c}
    \includegraphics{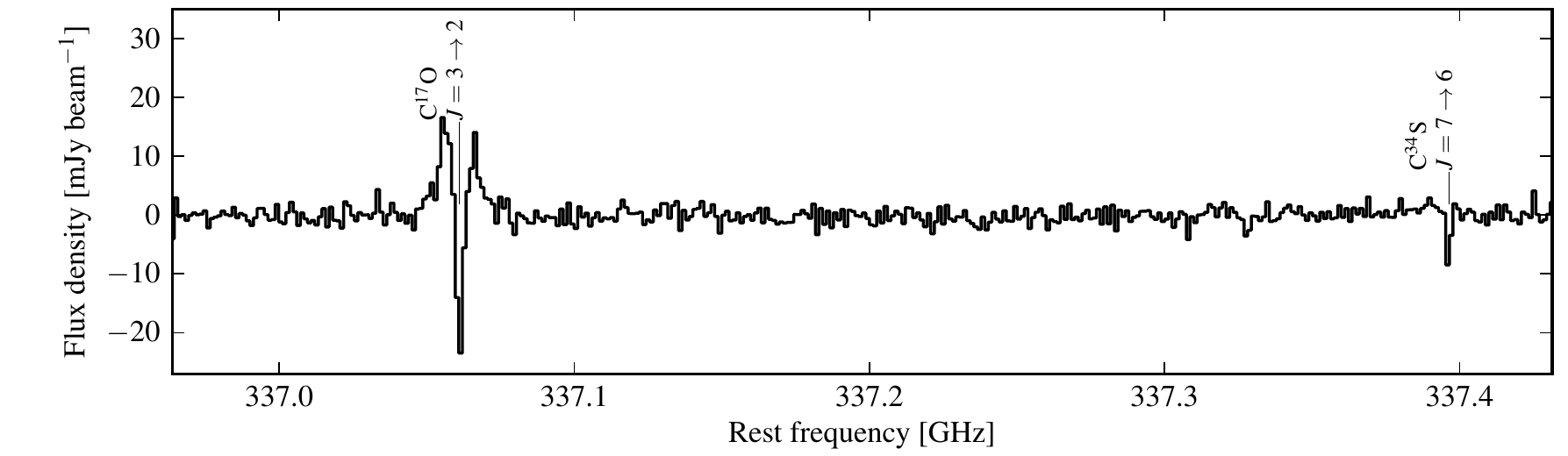} \\
	\includegraphics{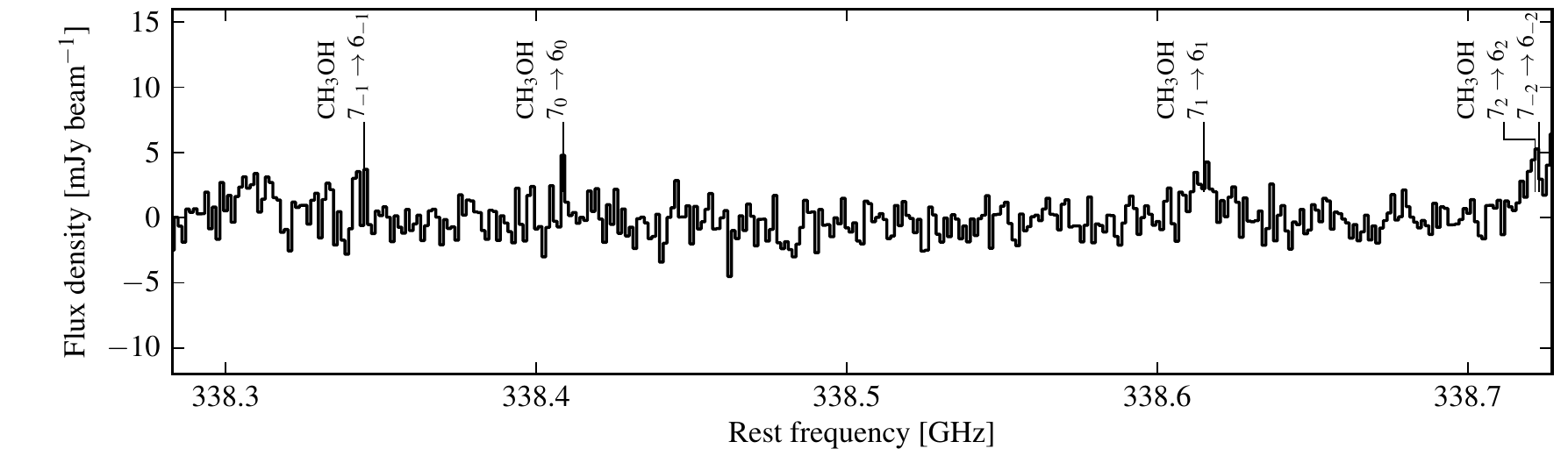} \\
	\includegraphics{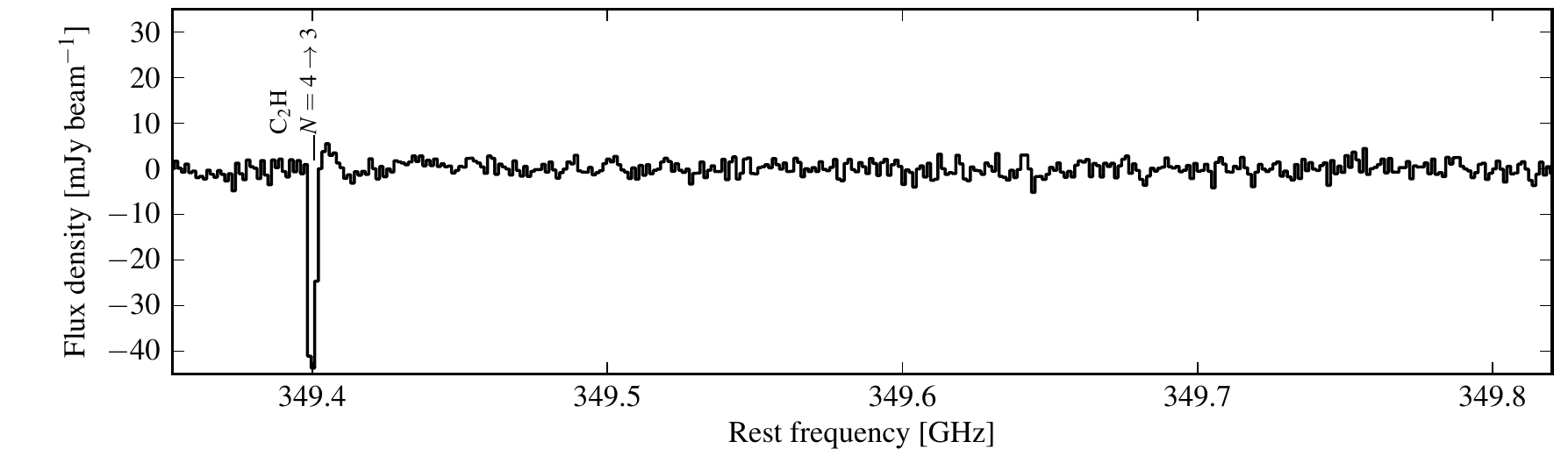} \\
	\includegraphics{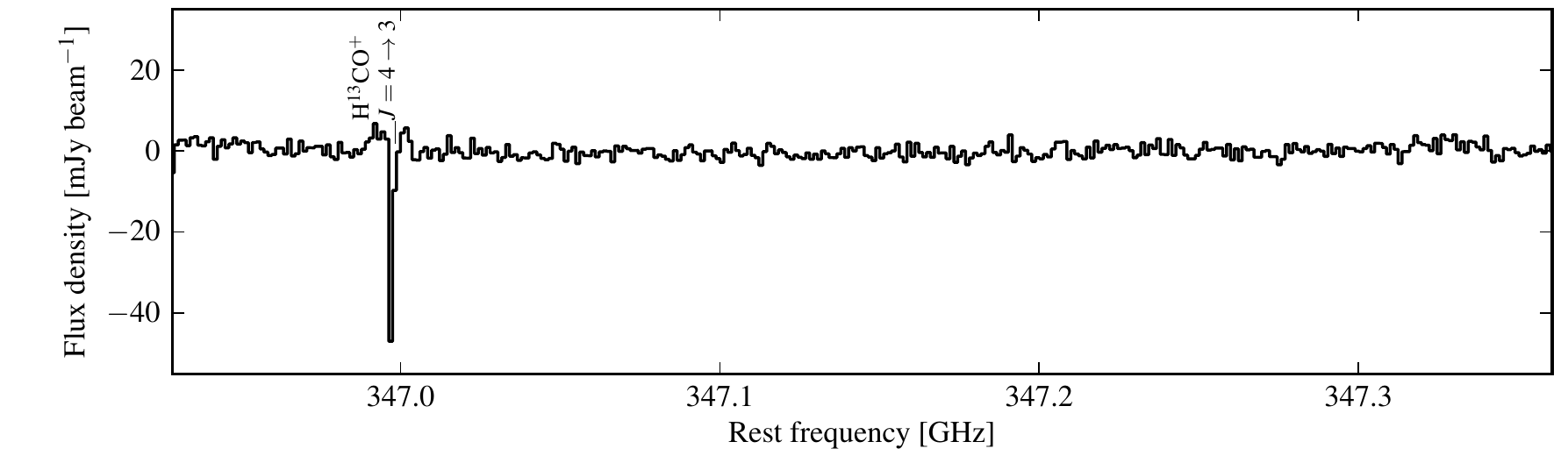} \\
    \end{array}$
    \caption{Full spectra of the four spectral windows averaged over a $1\farcs2\times1\farcs0$ box around IRS7B. The spectra are smoothed by a factor of 10 ($\sim1.1$~km~s$^{-1}$).}
    \label{fig:spectra1}
\end{figure*}

\begin{figure*}[!htb]
    \centering
	\includegraphics{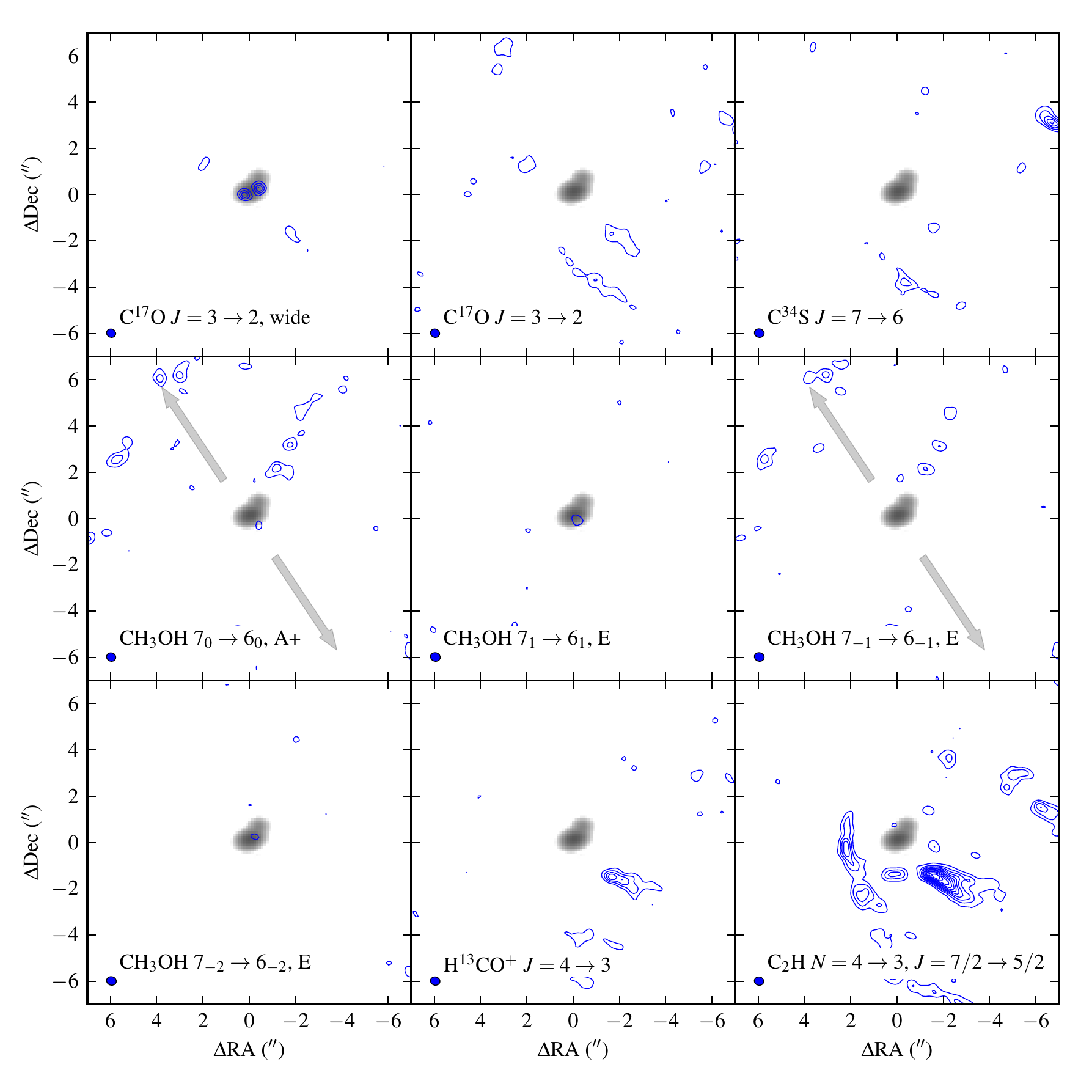}
    \caption{Moment 0 maps (intensity; blue contours) around IRS7B of all detected spectral lines. All lines are integrated between 4.0 and 8.8~km~s$^{-1}$, except for in the first C$^{17}$O map, where the line is integrated between $-1.7$ and $+13.3$~km~s$^{-1}$, to cover the high-velocity component of the emission. The contours are in steps of $3\sigma$, between 4.4 and 6.7~mJy~beam$^{-1}$~km~s$^{-1}$ (10.7~mJy~beam$^{-1}$~km~s$^{-1}$ for the wide C$^{17}$O map). The dust continuum emission is shown in greyscale. The large-scale CH$_3$OH outflow direction is shown with grey arrows.}
    \label{fig:mom0a}
\end{figure*}

Spectral lines from five different molecular species (C$^{17}$O, C$^{34}$S, CH$_3$OH, H$^{13}$CO$^+$, and C$_2$H) were detected in the ALMA data. The details of the detected transitions are shown in Table~\ref{tab:detections}. We will focus primarily on the spectra of IRS7B, the source at the phase centre of the observations (observed spectra towards this source are shown in Fig.~\ref{fig:spectra1}). Moment~0 maps of all detected lines in the region around IRS7B can be found in Fig.~\ref{fig:mom0a}.

Several of the molecular lines are detected as dips in the spectra at the systemic velocity towards IRS7B. This is indicative of extended emission being resolved out in the interferometric observations. This is consistent with the extended CH$_3$OH and H$_2$CO emission found by \citet{lindberg12}, and also \textit{Herschel} observations of CO, OH, and H$_2$O \citep{lindberg13}. Single-dish observations of the region performed with the ASTE telescope show very strong CH$_3$OH emission \citep{watanabe12}, with a peak flux density at $\sim25$~Jy for the $7_{0}\rightarrow 6_{0}$,~E line. In the ALMA data, the flux density of this line spatially integrated over the ALMA primary beam is $\sim1$~Jy, which shows that a few percent of the single-dish flux is recovered in the ALMA observations.

The major part of the CH$_3$OH emission is seen as a pattern of stripes and blobs in a NE/SW direction from IRS7B, which most likely are traces of an extended outflow resolved out in the interferometric data. In particular, this is seen in the lowest-energy CH$_3$OH transition $7_{0}\rightarrow6_{0}$,~E. This is consistent with a CH$_3$OH outflow seen in SEST data of the $2_{-2}\rightarrow1_{-1}$,~E and $2_{0}\rightarrow1_{0}$,~A+ transitions at 96.7~GHz (Miettinen et al., in prep.), extending northeast from IRS7B on scales of $\sim1$\arcmin. It is also consistent with the SMA/APEX observations of the CH$_3$OH $4_{2}\rightarrow 3_{1}$,~E line at 218~GHz, where strong extended emission was found towards SMM~1A south of IRS7B, but no significant emission was found towards the point sources in the field \citep{lindberg12}. Faint C$^{17}$O emission is also seen along roughly the same axis (see Fig.~\ref{fig:c17o_moments}), and the outflow direction is also consistent with recent observations of bipolar radio jets by \citet{liu14}. Faint on-source line emission consistent with four CH$_3$OH spectral lines is detected towards IRS7B. The significance of the four detections are between $3\sigma$ and $10\sigma$, which makes this a likely on-source detection of CH$_3$OH.

No complex organic molecules other than CH$_3$OH (including CH$_3$OCH$_3$ and CH$_3$CN that have known lines at 347.225~GHz and 349.454~GHz, respectively) were detected. The SiO $J=8\rightarrow7$ transition at 347.331~GHz is also not detected.

\begin{table}[!tb]
\centering
\caption[]{Detected spectral lines.}
\label{tab:detections}
\begin{tabular}{l l l l}
\noalign{\smallskip}
\hline
\hline
\noalign{\smallskip}
Species & Transition & Frequency\tablefootmark{a} & $E_{\mathrm{u}}$\tablefootmark{a} \\
& & [GHz] & [K] \\
\noalign{\smallskip}
\hline
\noalign{\smallskip}
C$^{17}$O & $J=3\rightarrow 2$ & 337.06112 & 32.353 \\
C$^{34}$S & $J=7\rightarrow 6$ & 337.39646 & 50.231 \\
CH$_3$OH & $7_{-1}\rightarrow 6_{-1}$,~E & 338.34463 & 70.550 \\
CH$_3$OH & $7_{0}\rightarrow 6_{0}$,~A+ & 338.40868 & 64.981 \\
CH$_3$OH & $7_{1}\rightarrow 6_{1}$,~E & 338.61500 & 86.051 \\
CH$_3$OH & $7_{2}\rightarrow 6_{2}$,~E & 338.72163 & 87.258 \\
CH$_3$OH & $7_{-2}\rightarrow 6_{-2}$,~E & 338.72294 & 90.912 \\
H$^{13}$CO$^+$ & $J=4\rightarrow 3$ & 346.99834 & 41.634 \\
C$_2$H & $N=4\rightarrow3$\tablefootmark{b} & 349.39927 & 41.927 \\
C$_2$H & $N=4\rightarrow3$\tablefootmark{c} & 349.40067 & 41.928 \\
\noalign{\smallskip}
\hline
\end{tabular}
\tablefoot{
	\tablefoottext{a}{From the CDMS database \citep{cdms}.}
	\tablefoottext{b}{$J=7/2\rightarrow5/2, F=4\rightarrow3$}
	\tablefoottext{c}{$J=7/2\rightarrow5/2, F=3\rightarrow2$}
		
     	}
\end{table}

\section{Analysis}

\subsection{A Keplerian disc around IRS7B}
\label{sec:analysis_keplerian}

\begin{figure}[!htb]
    \centering
    \includegraphics{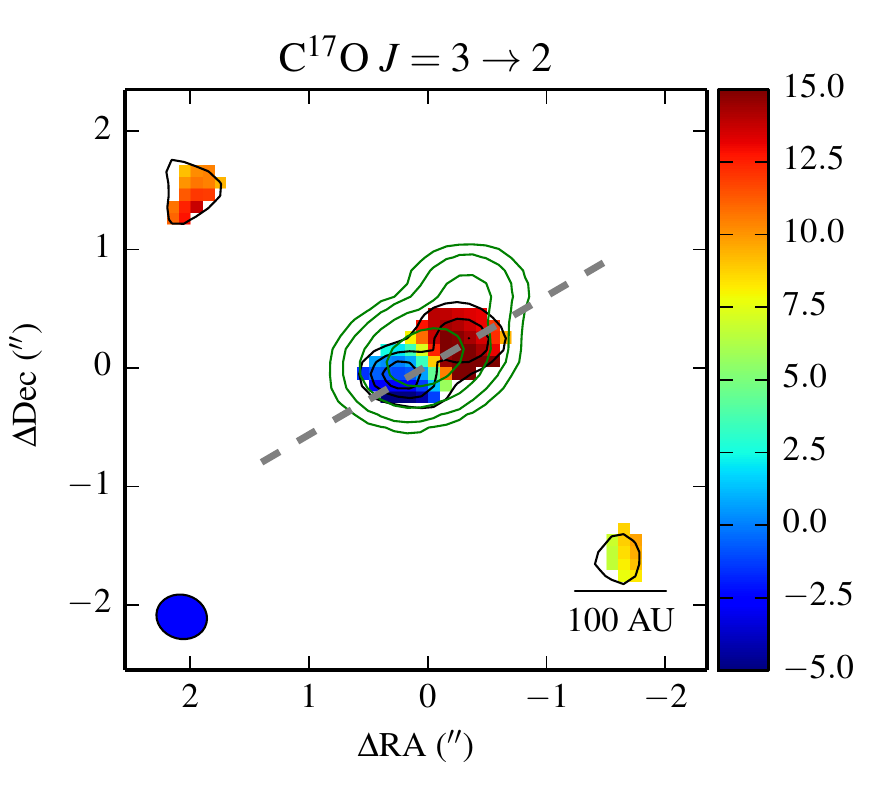}
    \caption{Moment 0 map (black contours at $3\sigma \approx 43$~mJy~beam$^{-1}$~km~s$^{-1}$ intervals) and moment 1 map (colour scale, velocities in km~s$^{-1}$) of the C$^{17}$O emission centred at IRS7B, integrated between $-11$~km~s$^{-1}$ and $+21$~km~s$^{-1}$. The colour bar indicates LSR velocities in km~s$^{-1}$. The dashed grey line indicates the axis along which the PV diagram fits of Figs.~\ref{fig:pvdiag3}--\ref{fig:pvdiag2} are made. The dust continuum emission is shown in green contours (logarithmically spaced between 2~mJy and 200~mJy).}
    \label{fig:c17o_moments}
\end{figure}

\begin{figure}[!htb]
    \centering
    \includegraphics{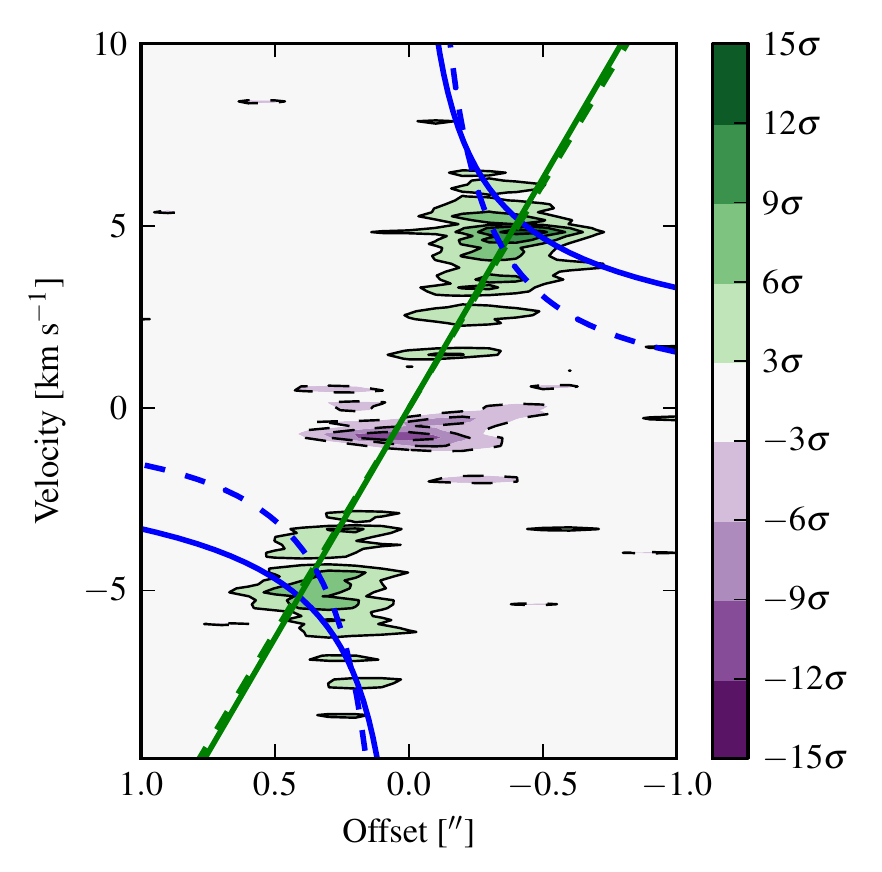}
    \caption{Position-velocity (PV) diagram of the cleaned image data of the C$^{17}$O emission centred at IRS7B. The velocities are relative to the centroid velocity, which has a $\varv_{\mathrm{LSR}} = 6.2$~km~s$^{-1}$. The solid lines show the best $\chi^2$-test fits to Keplerian rotation ($\varv\propto r^{-1/2}$ component in blue and $\varv\propto r$ component in green; corresponding to a central mass of $M_{\star}=2.3~M_{\odot}$ assuming the most probable inclination of 60\degr). The dashed lines show the best $\chi^2$-test fits to infall under conservation of angular momentum ($\varv\propto r^{-1}$ component in blue and $\varv\propto r$ component in green).}
    \label{fig:pvdiag3}
\end{figure}

\begin{figure}[!htb]
    \centering
    \includegraphics[width=1.0\linewidth]{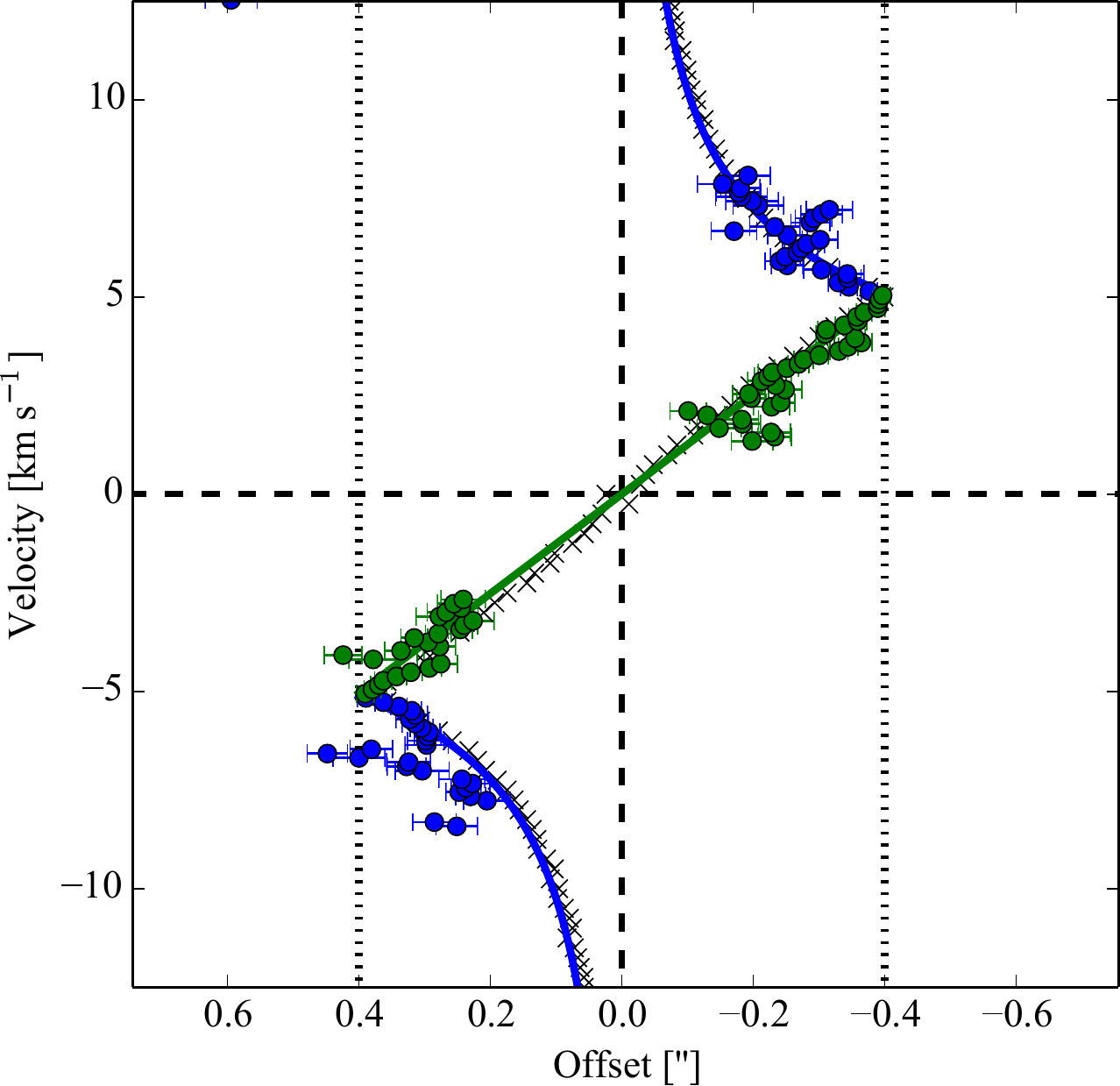}
    \caption{Position-velocity (PV) diagram of the \texttt{uvmodelfit} peaks in the C$^{17}$O emission centred at IRS7B (blue and green data points). The black crosses are the result of an LTE calculation of a simple flat, Keplerian disc with a radius marked by the vertical, dotted lines. The solid lines as in Fig.~\ref{fig:pvdiag3}.}
    \label{fig:newplot}
\end{figure}

\begin{figure*}[!htb]
    \centering
    \centering
    $\begin{array}{c@{\hspace{0.0cm}}c}
    \includegraphics{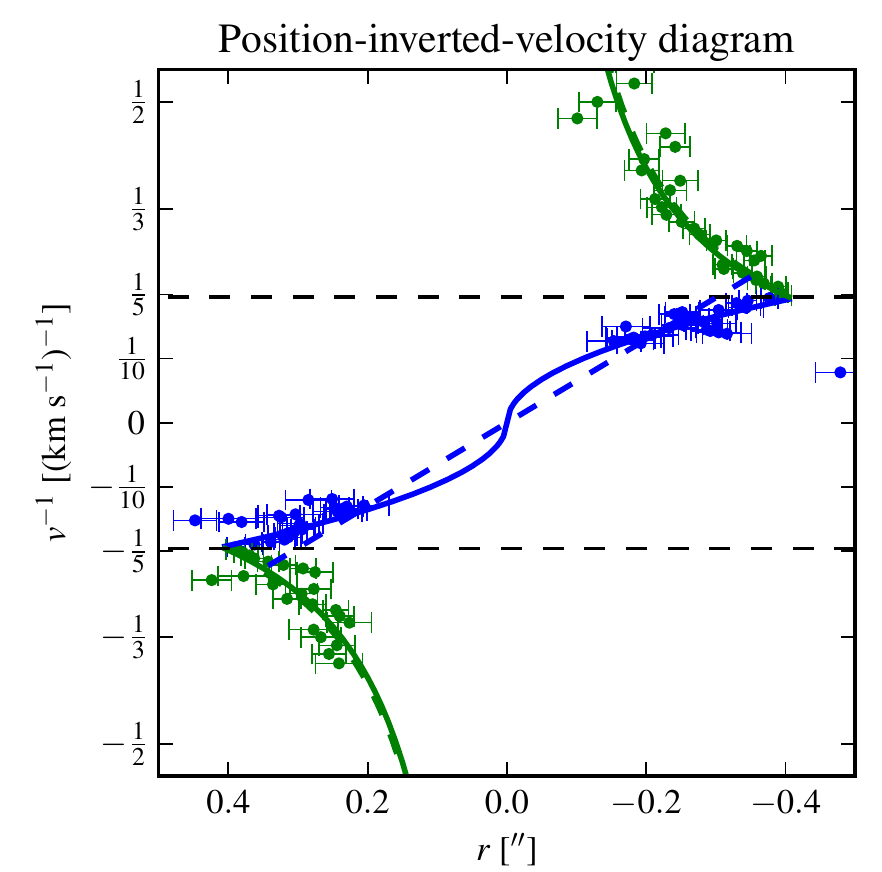}
    \includegraphics{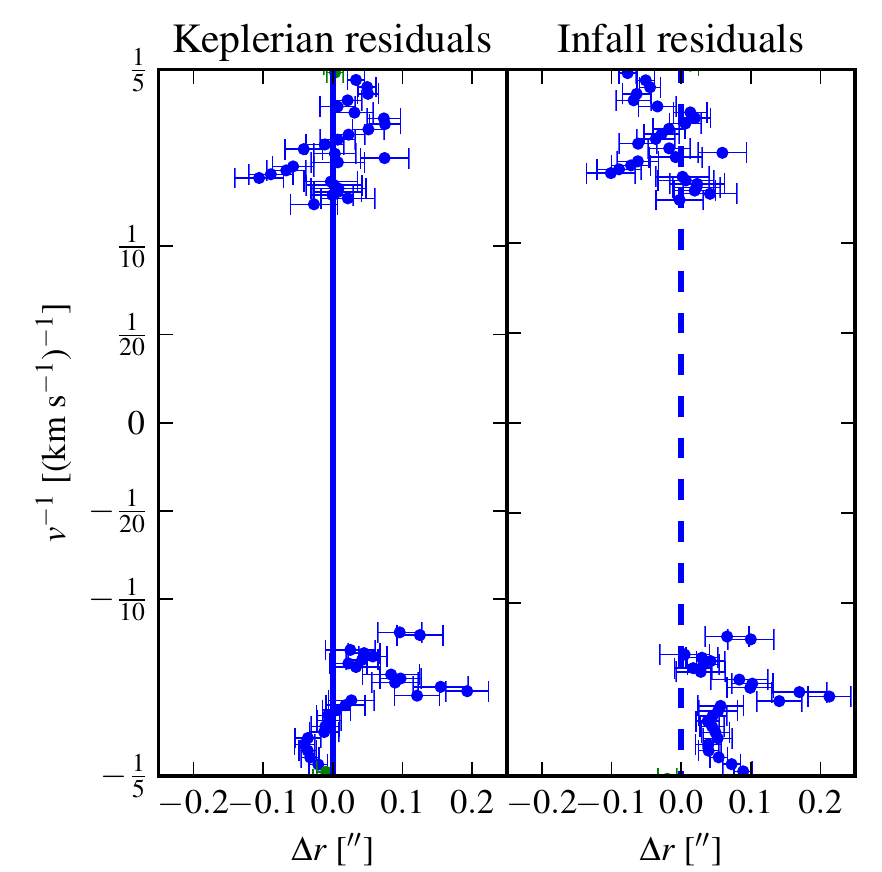}
    \end{array}$
    \caption{\textit{Left:} Position-inverted-velocity (PIV) diagram of the \texttt{uvmodelfit} peaks in the C$^{17}$O emission centred at IRS7B. The inverted velocity axis is used to make the origin of the diagram correspond to the central object. Blue dots are used for the $\varv\propto r^{-1/2}$ and $\varv\propto r^{-1}$ fits, green dots for the $\varv\propto r$ fit. The velocities are given relative to the centroid velocity, which has a $\varv_{\mathrm{LSR}} = 6.2$~km~s$^{-1}$. The solid and dashes lines as in Fig.~\ref{fig:pvdiag3}.
    \textit{Right:} Position residuals of the blue (high velocity) data points plotted versus inverse velocity for the two models (Keplerian rotation and infall under conservation of angular momentum), zoomed in to the dashed part of the left panel.
    }
    \label{fig:pvdiag2}
\end{figure*}

In the C$^{17}$O $J=3\rightarrow 2$ data we find strong point-like
emission around IRS7B with a large velocity span. The moment~0 
  (integrated intensity) and moment~1 (velocity) maps of this
line are overplotted in Fig.~\ref{fig:c17o_moments}. Through the CASA
task \texttt{uvmodelfit}\footnote{\url{http://casa.nrao.edu/docs/taskref/uvmodelfit-task.html}} we identify the strongest peak of the
C$^{17}$O emission in each channel of the $(u,\varv)$ data. By the method
of \citet{goodman93}, which uses a weighted least-square fit to the positional data to establish the inclination of the velocity gradient on the plane of sky, we construct a
position-velocity diagram (PV diagram) of the C$^{17}$O emission,
shown in Figs.~\ref{fig:pvdiag3}--\ref{fig:pvdiag2}. We find that the
motion has a position angle of $-65$\degr\ on the sky and is centred
around the channel corresponding to an LSR velocity of
6.2~km~s$^{-1}$. The velocities in the PV diagram are given relative
to this central velocity.

\begin{figure}[!htb]
    \centering
    \includegraphics{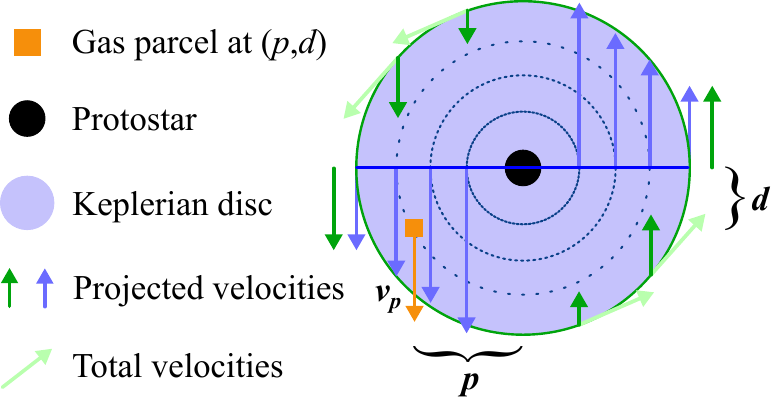}
    \caption{Schematic illustration of the disc explaining the green and blue velocity components in Fig.~\ref{fig:newplot}. Radial velocities as seen by an observer below the image. The blue and green arrows correspond to the blue and green data points in Figs.~\ref{fig:newplot}--\ref{fig:pvdiag2}. The illustration is not drawn to scale.}
    \label{fig:sketch}
\end{figure}

The data in the PV diagram (Fig.~\ref{fig:newplot}) are consistent with a simple flat disc model with Keplerian rotation. The data can be separated into two parts; one part that lies beyond $\pm5.1$~km~s$^{-1}$ which is roughly distributed like a $r^{-0.5}$ profile and one part within $\pm 5.1$~km~s$^{-1}$ which roughly follows a linear distribution. These two parts have been coloured blue and green, respectively, in Figs.~\ref{fig:pvdiag3}--\ref{fig:pvdiag2}. Shown with black crosses in Fig.~\ref{fig:newplot} is a simple fiducial disc model for which the C$^{17}$O line emission has been calculated under the assumption of LTE using the LIME code \citep{brinch10}. The model data points have been extracted directly from the simulated image. The physical structure of the disc is unimportant as long as the emission is optically thin, because we are only interested in the kinematic signature. The central mass and the disc radius are the only free parameters. In the model, the disc is assumed to be flat with a radial density profile $n(r) \propto r^{-1}$ and a temperature profile $T(r) \propto r^{-0.5}$.

The split in the distribution of the data points into these two parts is what is expected for a Keplerian disc with a finite radius as can be seen from simple considerations of an optically thin line emission: the line-of-sight velocity $\varv_p$ in a Keplerian disc seen edge-on at a projected distance of $p$ and at a depth $d$ (see Fig.~\ref{fig:sketch}) can be written as
\begin{equation}
\varv_p = \frac{(G M_{\star})^{1/2}}{(p^2 + d^2)^{1/4}} \frac{p}{(p^2 + d^2)^{1/2}}\,.
\end{equation}
The flux in a given velocity channel $\Delta \varv_p$ at a given projected distance $p$ is then
\begin{equation}
F(\Delta \varv_p,p) = \int_{\Delta \varv_p} \epsilon\,d(\varv_p,p)\,d\varv_p\,,
\end{equation}
where the emission factor $\epsilon$ is proportional to the flux from a unit volume, and $d(\varv_p,p)$ for a given projected distance $p$ is the line-of-sight distance at which the disc will have a projected velocity of $\varv_p$:
\begin{equation}
d(\varv_p,p) = \left[ (G M_{\star})^{2/3} p^{4/3} \varv_p^{-4/3} - p^2\right]^{1/2}\,.
\end{equation}
We assume that the disc is optically thin, in which case $\epsilon
\propto \rho T$, and that $\epsilon \propto r^{-\gamma}$ with $\gamma
< 5/2$. In this case, for projected velocities above the Keplerian
velocity at the edge of the disc, the maximum flux is along a radial
line in the plane of the sky giving rise to the $\varv_p \propto \pm
p^{-1/2}$ branches (blue lines in
Figs.~\ref{fig:pvdiag3}--\ref{fig:pvdiag2}). For projected velocities
below the Keplerian velocity at the edge of the disc, the maximum flux
is at the edge of the disc. If the disc radius is $r_{\mathrm{D}}$
then $\varv_p = (G M_{\star})^{1/2} r_{\mathrm{D}}^{-3/4} p$ along the
edge of the disc, giving a linear slope in the PV diagram (green lines
in Figs.~\ref{fig:pvdiag3}--\ref{fig:pvdiag2}). For a disc that is not
edge-on, but at an inclination, the usual geometric projection factors
have to be applied.

The disc edge falls exactly where the two parts join (marked by the two dotted lines in Fig.~\ref{fig:newplot}), and by fitting a Keplerian profile (blue curve) and a straight line (green curve) to the data we can determine the disc radius very accurately. We find the radius of the spatially well-constrained disc to be 0.41$''$ which equals to 50~AU at the distance of IRS7B. The best-fit Keplerian profile corresponds to a central mass of $M_{\star}=2.3~M_{\odot}$ assuming the most probable inclination of $60\degr$.

Note that future, higher quality, data probing the detailed shape of the low velocity branches (green lines in Figs.~\ref{fig:newplot} and \ref{fig:sketch}) may potentially give direct information on $\gamma$, i.e. the thermal energy profile of the disc. Specifically, depending on the exact value of $\gamma$, below a critical velocity in the PV diagram, the maximum flux will be close to the centre instead of at the edge of the disc. The transition point depends directly on $\gamma$.

We also explored to what extent the data in the PV diagram can be described by the emission that is expected from material falling in under angular momentum conservation ($\varv\sim r^{-1}$; \citealt{yen11,belloche13}). We do a $\chi^2$ fit to the velocities as a function of radius weighted by the \texttt{uvmodelfit} errors on the positional fits. The fits are made for a combination of a $\varv\sim r$ profile for the lower velocities in both fits, and $\varv\sim r^{-1/2}$ for the Keplerian fit and $\varv\sim r^{-1}$ for the infall fit. The free parameters are the disc size and the central mass. We find that the two fits have similar $\chi^2$ values (the Keplerian fit has a $\chi^2$ value a factor 2.3 times lower than the infall fit), and thus neither fit can be completely ruled out. However, since the $\varv\sim r$ profile arises from the edge of the disc, this is an unlikely effect in the infall scenario, since the infalling material should not have a sharp edge, as observed. In Figs.~\ref{fig:pvdiag3} and \ref{fig:pvdiag2}, the Keplerian fit is shown with solid lines and the infall fit with dashed lines. In the right panel of Fig.~\ref{fig:pvdiag2}, the residuals of the high-velocity data points are compared for the two models. It can be seen that the Keplerian model gives a better fit, in particular for the lower velocities, but the model with infall under conservation of angular momentum cannot be excluded, if the well-defined edge is disregarded. These relatively simple models provide good fits to the data, but because of the high sensitivity of the data, they may still contain traces of features which are not explained by these models.

Assuming that the continuum emission originates in a disc, we can estimate the mass of the disc around IRS7B from the ALMA continuum emission using the method of \citet{jorgensen07} assuming optically thin emission and a dust temperature of 30~K (see Table~\ref{tab:pointsource}). The peak flux of the 0.8~mm continuum emission is 432~mJy, which corresponds to a disc mass of $0.024~M_{\odot}$, or 1.0\% of the central source in the Keplerian disc ($M_{\star}=2.3~M_{\odot}$), and 1.1\% of the envelope mass ($M_{\mathrm{env}}=2.2~M_{\odot}$; see below). This is relatively low compared to the disc/star mass ratio in Class~I objects and disc/envelope mass ratios in Class~0 and Class~I objects \citep{jorgensen09}.

Considering the small angular size of the three remaining point sources SMM~1C, CXO~34, and IRS7A, the continuum emission at these positions is most likely dust emission from discs. Except for faint CH$_3$OH emission at the position of IRS7A, no molecular emission is detected towards these point sources, which makes it impossible to study their kinematics. However, they are situated on the edge of or beyond the primary beam of the observations, so the non-detections of molecular line emission in these sources are not significant.

\subsection{CH$_3$OH modelling}

\begin{figure*}[!htb]
    \centering
    \includegraphics{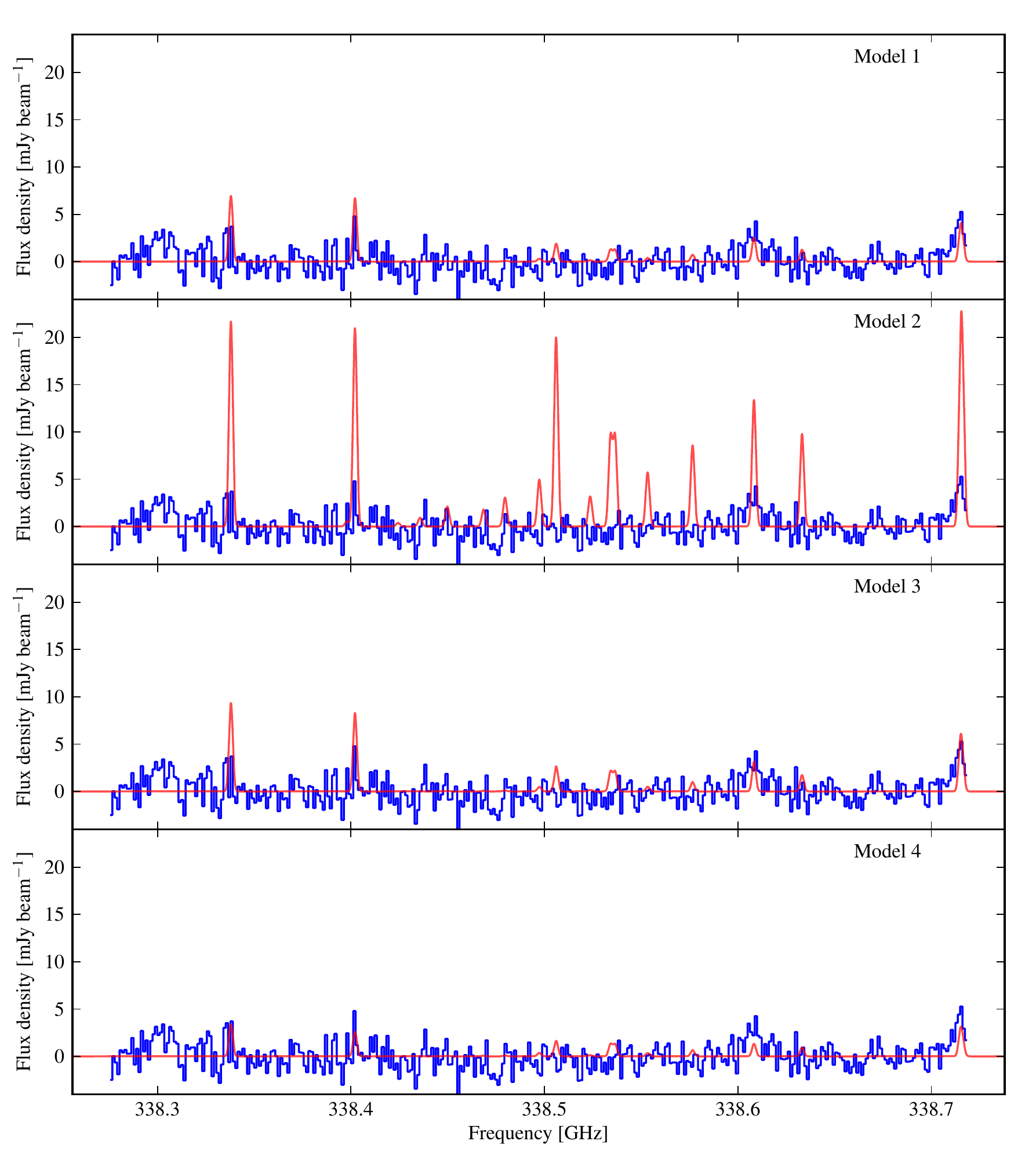}
    \caption{RATRAN models of CH$_3$OH (red) and IRS7B ALMA spectral window 1 spectrum smoothed to 1.1~\kms\ channels (blue). The density profile follows a $-1.5$ power law at $r>100$~AU in all models. In Models 1--2 it follows this power law at all $r$. In Models~3--4 it is flat for $r<100$~AU. The CH$_3$OH abundance is constant at $10^{-10}$ in Model~1. In the remaining models, it has a break profile at $T=100$~K, with an inner CH$_3$OH abundance of $10^{-8}$. The outer CH$_3$OH abundance is $10^{-10}$ in Models~2--3 and $10^{-12}$ in Model~4. The model parameters are also found in Table~\ref{tab:models}.}
    \label{fig:ch3oh_model}
\end{figure*}

The CH$_3$OH emission in a $1\farcs2\times1\farcs0$ box around IRS7B was modelled using the radiative transfer modelling tool RATRAN \citep{ratran}. CH$_3$OH collision rates from the LAMDA database were used \citep{lamda,rabli10}. The level population files produced after convergence of each model were examined, and with the exception of some of the highest levels ($E_{\mathrm{u}} > 600$~K; with fractional populations $\lesssim10^{-18}$), the level populations as function of radius were found to be continuous and well-behaved. All the model spectra are shown together with the observed CH$_3$OH spectrum in Fig.~\ref{fig:ch3oh_model}, and the parameters used in each model are summarised in Table~\ref{tab:models}.

\begin{table}[!tb]
\centering
\caption[]{CH$_3$OH RATRAN model parameters.}
\label{tab:models}
\begin{tabular}{l l l l}
\noalign{\smallskip}
\hline
\hline
\noalign{\smallskip}
Model \# & $n_{<100~\mathrm{AU}}$\tablefootmark{a} profile & $X_{\mathrm{inner}}$\tablefootmark{b} & $X_{\mathrm{outer}}$\tablefootmark{c} \\
\noalign{\smallskip}
\hline
\noalign{\smallskip}
1 & $n \propto r^{-1.5}$ & $10^{-10}$ & $10^{-10}$ \\
2 & $n \propto r^{-1.5}$ & $10^{-8}$ & $10^{-10}$ \\ 
3 & $n$ = constant & $10^{-8}$ & $10^{-10}$ \\ 
4 & $n$ = constant & $10^{-8}$ & $10^{-12}$ \\
\noalign{\smallskip}
\hline
\end{tabular}
\tablefoot{
	\tablefoottext{a}{Density profile at $r < 100$~AU. The density profiles at $r > 100$~AU follow $n \propto r^{-1.5}$ in all models.}
	\tablefoottext{b}{CH$_3$OH abundance profile in the hot inner envelope, $r < r_{100~\mathrm{K}}$ (where $T>100$~K).}
	\tablefoottext{c}{CH$_3$OH abundance profile at $r > r_{100~\mathrm{K}}$ (where $T<100$~K).}
     	}
\end{table}

The envelope model of \citet{lindberg12} was adopted. In this model the density profile follows an $r^{-1.5}$ power law with a mass of $2.2~M_{\odot}$. The temperature in the envelope is governed by heating from the central object on small scales ($<500$~AU), but by the external irradiation from R~CrA on large scales ($>1000$~AU). In Model 1, a constant CH$_3$OH abundance was assumed. The best fit was achieved for a CH$_3$OH abundance of $10^{-10}$, which is much lower than expected in a source with a hot inner envelope. However, IRS7B is one of the youngest sources exhibiting X-ray emission \citep{hamaguchi05}, which possibly could be responsible for the destruction of CH$_3$OH. The external irradiation from R~CrA could also be responsible for a chemistry with less complex organic molecules through the early evaporation of CO from the icy grain mantles \citep{lindberg12}.

In the remaining models we assumed a jump abundance of CH$_3$OH as proposed by \citet{schoier02,jorgensen05,maret05}. This model is used since the evaporation of CH$_3$OH from grains is enhanced by at least two orders of magnitudes at the ice sublimation temperature $T \sim 100$~K \citep[][and references therein]{herbst09}. The jump in the abundance profile is thus introduced at the radius of the hot inner envelope, $r_{100~\mathrm{K}} \approx 30$~AU, inside which $T>100$~K. This radius is fairly insensitive to modifications of the envelope model and central luminosity -- if the density profile power law is changed from $-1.5$ to $-2.0$, or if the central luminosity is changed by 20\%, $r_{100~\mathrm{K}}$ only changes by a few AU. With an inner CH$_3$OH abundance of $10^{-8}$ and an outer CH$_3$OH abundance of $10^{-10}$, the model CH$_3$OH line fluxes are much stronger than in the observed data. 

To account for the disc detected in the C$^{17}$O, we assume a flat density profile at $r<r_{\mathrm{cf}}$ where $r_{\mathrm{cf}} = 100$~AU is the assumed centrifugal radius. The density profile at $r>r_{\mathrm{cf}}$~AU and the shape of the abundance profile were kept as in Model 2 (power law density profile with a jump abundance). With the inner CH$_3$OH abundance set to $10^{-8}$ and the outer to $10^{-10}$ (Model 3), the observed fluxes are reproduced to a first-order approximation. Another model with the outer abundance lowered to $10^{-12}$ (Model 4) is also tested, but it produces fluxes that are only somewhat weaker than in the observations, meaning that the CH$_3$OH emission in these models is dominated by the compact component. Thus, Models 3 and 4 provide equally good fits to the data.

All models overestimate the non-detected CH$_3$OH lines at 338.51~GHz and 338.53~GHz. The small change in these line strengths between Models 3 and 4 show that these lines are tracing the inner envelope with just a small contribution from material at larger scales. The inner CH$_3$OH abundance of these models ($10^{-8}$) should thus be seen as an upper limit.

The faint on-source CH$_3$OH emission either puts a constraint
  on the CH$_3$OH abundance in the inner envelope to be as low as
  $\sim 10^{-10}$, or requires that the envelope has a flattened
  density profile at $R\lesssim 100$~AU. This contrasts to earlier
  single-dish studies of several other Class~0/I sources
  \citep{jorgensen05,maret05}, for which radiative transfer models
  showed CH$_3$OH abundances $\sim10^{-8}$--$10^{-7}$ in the innermost
  $T > 90-100$~K regions -- and is at least two orders of magnitude lower than the values or upper limits for all sources in the sample of \citet{maret05}.

\section{Discussion}

\citet{lindberg12} suggested that the high level of irradiation from R~CrA onto IRS7B would have a considerable effect on the chemistry in the envelope due to premature evaporation of CO and CH$_4$ from the dust grains. Irradiation from R~CrA in an early stage of the IRS7B core would prevent complex organic molecules and carbon-chain molecules to form on the dust grains. This together with photo-dissociation of H$_2$O, HCN, and other molecules would produce a chemistry dominated by PDR (photo-dissociation region) tracers such as OH and CN, and no sign of either hot inner envelope (``hot corino''; \citealt{ceccarelli05}) chemistry or warm carbon-chain chemistry \citep[WCCC;][]{sakai09a,sakai09b}, in agreement with single-dish line-survey observations \citep[][Lindberg et~al., in~prep.]{watanabe12} and \textit{Herschel} PACS spectrometer observations \citep{lindberg13}. 

However, H$_2$CO \citep{lindberg12} and CH$_3$OH emission is predominantly found on large scales in the protostellar envelope. Only very faint CH$_3$OH emission is found directly towards the IRS7B continuum source position in the ALMA data.

The C$^{17}$O data indicate the presence of a disc with a radius on 50~AU scales, which could be responsible for a flattened density profile or an inner edge of the envelope at $r\sim50$--$100$~AU, which is supported by the faint CH$_3$OH lines towards the continuum source and the results of the modelling of the CH$_3$OH line observations.

Assuming an optically thin envelope heated only by a central source, the dust temperature distribution in the envelope follows
\begin{equation}
T_{\mathrm{d}}(r) \sim r^{-q} L_{\star}^{q/2}
\label{eq:temp_r}
\end{equation}

\noindent
where $q$ relates to the power-law index $\beta$ of the dust-opacity law as $q = 2/(4 + \beta)$ (\citealt{jorgensen06} and references therein). As in the previous section, we define the radius of the hot inner envelope, $r_{100~\mathrm{K}}$, as the radius within which the temperature exceeds 100~K, allowing for the evaporation of CH$_3$OH and other complex organic molecules from the grains. Using Eq.~\ref{eq:temp_r}, we find that
\begin{equation}
r_{100~\mathrm{K}} \sim L_{\star}^{1/2},
\end{equation}

\noindent
independently of the value of $q$. If $L_{\star}$ is dominated by accretion ($L_{\mathrm{i}} \sim GM_{\star}\dot{M_{\mathrm{i}}}/r_{\star}$), it scales linearly with $M_{\star}$, and 
\begin{equation}
r_{100~\mathrm{K}} \sim M_{\star}^{1/2}.
\end{equation}

The size of the disc is on the other hand characterised by the centrifugal radius \citep{terebey84}, at which the rotational support gets greater than the gas pressure support due to the increase in angular momentum:
\begin{equation}
r_{\mathrm{cf}} \approx 0.5 \frac{GM_{\star}}{c_{\mathrm{s}}^2}.
\end{equation}

The hot inner envelope may be larger than the disc at early times. However, as its size follows $r_{100~\mathrm{K}} \sim M_{\star}^{1/2}$ compared to the disc size $r_{\mathrm{cf}} \sim M_{\star}$, the chemistry on small scales will at later times be dominated by the disc.

IRS7B shows relatively low CH$_3$OH abundances and non-detections of other complex organic molecules, but also a disc at the same or slightly larger scales as the hot inner envelope would be expected to be present at. We propose that it has entered the stage where the disc dominates the physical and chemical structure at small radii.

\section{Conclusions}

We have performed high-resolution spectral line and continuum observations of four low-mass Class~0/I protostars in the R~CrA cloud using the ALMA interferometer. The observations show the structure of the molecular gas and the dust emission at resolutions $\sim50$--$1000$~AU. Our most important findings are: 

   \begin{enumerate}
      \item The C$^{17}$O emission around the Class~0/I protostar R~CrA IRS7B is indicative of a Keplerian disc in C$^{17}$O emission, consistent with a disc mass $M_{\mathrm{disc}}\approx0.024~M_{\odot}$ and a central source mass $M_{\star}\approx2.3~M_{\odot}$. The combined mass of the disc and star is thus similar to the mass of the envelope ($M_{\mathrm{env}}=2.2~M_{\odot}$; \citealt{lindberg12}), which implies a source at a very young stage. The radius of the disc can be accurately determined to 50~AU. A scenario where the observed C$^{17}$O emission rises from material falling in under conservation of angular momentum can, however, not be excluded, but seems unlikely due to the well-defined edge of the disc which is observed.
      
      \item Outflow-like structures are detected around IRS7B on large ($>1000$~AU) scales in CH$_3$OH emission and on smaller ($\sim500$~AU) scales in faint C$^{17}$O emission, roughly perpendicular to the C$^{17}$O disc emission.
      
      \item CH$_3$OH emission is found on large scales ($>1000$~AU) in the ALMA data, but is only marginally detected on the IRS7B continuum point source itself. The observed CH$_3$OH spectral line intensities at IRS7B are consistent with radiative transfer models assuming an envelope with a flat density profile inside 100~AU and CH$_3$OH abundances of $10^{-8}$ in the hot inner envelope (where $T>100$~K) and substantially lower outside this region.
      
      \item We propose that the chemistry of IRS7B on small scales is dominated by the presence of the disc -- which is more extended than the region of the envelope where the temperature is greater than 100~K. The presence of discs around typical Class~I young stars may cause these sources not to show significant emission from complex organic molecules characteristic of such hot inner envelopes.
      
   \end{enumerate}
   
To verify that the C$^{17}$O emission on IRS7B originates in a Keplerian disc, higher-sensitivity and higher-resolution observations are needed. Such observations would also be useful to investigate whether complex organic molecules are present at all in IRS7B on the scales of a hot inner envelope ($\lesssim 30$~AU).

\begin{acknowledgements}
  The authors would like to thank Iv\'{a}n Mart\'{i}-Vidal and Wouter Vlemmings at the Nordic ALMA Regional Centre node at Onsala Space Observatory for support with the calibration and data
  reduction. This paper makes use of the following ALMA data:
  ADS/JAO.ALMA\#2011.0.00628.S. ALMA is a partnership of ESO
  (representing its member states), NSF (USA) and NINS (Japan),
  together with NRC (Canada) and NSC and ASIAA (Taiwan), in
  cooperation with the Republic of Chile. The Joint ALMA Observatory
  is operated by ESO, AUI/NRAO and NAOJ. Research at Centre for Star
  and Planet Formation is funded by the Danish National Research
  Foundation and the University of Copenhagen's programme of
  excellence. This research was also supported by a Lundbeck
  Foundation Group Leader Fellowship to JKJ. TH is supported by a
  Sapere Aude Starting Grant from The Danish Council for Independent
  Research. DH is supported by the Netherlands Research School for
  Astronomy (NOVA) and by the Space Research Organization Netherlands
  (SRON). MVP acknowledges EU FP7 grant 291141 CHEMPLAN.
\end{acknowledgements}

\bibliographystyle{aa}
\bibliography{almapaper_accepted}
\clearpage
\onecolumn

\end{document}